\documentclass[a4paper]{article}
\usepackage{multirow}
\usepackage{subfig}
\usepackage{graphicx}
\usepackage{INTERSPEECH2022}
\usepackage{amsmath}
\usepackage{amssymb}
\usepackage{pifont}
\usepackage{arydshln}
\usepackage{hyperref}
\usepackage{arydshln}

\usepackage{bbm}

\usepackage{float}
\usepackage[disable]{todonotes}

\hypersetup{
    colorlinks=true,
    linkcolor=blue,
    filecolor=magenta,      
    urlcolor=blue,
    pdftitle={Overleaf Example},
    pdfpagemode=FullScreen,
}
\title{Robustness of Neural Architectures for Audio Event Detection}
\name{Juncheng B Li$^1$, Zheng Wang, Shuhui Qu, Florian Metze$^2$}
\address{Carnegie Mellon University}
\email{junchenl@cs.cmu.edu}

\begin{document}
\maketitle
\begin{abstract}
Traditionally, in Audio Recognition pipeline, noise are suppressed by the ``frontend", relying on preprocessing techniques such as speech enhancement.
However, it is not guaranteed that noise will not cascade into downstream pipelines.
To understand the actual influence of noise on the entire audio pipeline, in this paper, we directly investigate the impact of noise on different type of neural models without the preprocessing step. 
We measure the recognition performances of 4 different neural network models on the task of environment sound classification under the 3 types of noises: \emph{occlusion} (to emulate intermittent noise), \emph{Gaussian} noise (models continuous noise), and \emph{adversarial perturbations} (worst case scenario).
Our intuition is that the different ways in which these models process their input (i.e. CNNs have strong locality inductive biases, which Transformers do not have) should lead to observable differences in performance and/ or robustness, an understanding of which will enable further improvements. 
We perform extensive experiments on AudioSet~\cite{gemmeke2017audio} which is the largest weakly-labeled sound event dataset available.
We also seek to explain the behaviors of different models through output distribution change and weight visualization.\footnote{Preprint Mar 2022}
\todo{F: I think we can clean this up a bit: the abstract does not need to repeat the introduction ("noise robustness is essential ..."), it should just say what we do (we compare three different neural architectures for audio classification with respect to their robustness against various types of noise), maybe with a bit of a motivations, and summarize the findings.}

\end{abstract}
\todo{F: adversarial techniques?}

\section{Introduction}
As various audio recognition (AR) systems such as Automatic Speech Recognition (ASR), acoustic event detection (AED) are being deployed onto our phones; into cars; indoor \& outdoor; \emph{noise robustness} is becoming ever more crucial to any type of AR system~\cite{li2019adversarial}.
Traditional AR pipeline relies on preprocessing/enhancement step to filter out the noise that exists in the input. 
However, enhancement by no means guarantees noise free~\cite{michelsanti2021overview, wang2018supervised}, noise still cascades to downstream pipelines. Therefore, it is also important to understand the robustness of models. 
\begin{figure}[t]
\hskip-0.8cm\centering
    \includegraphics[width=1.1\linewidth]{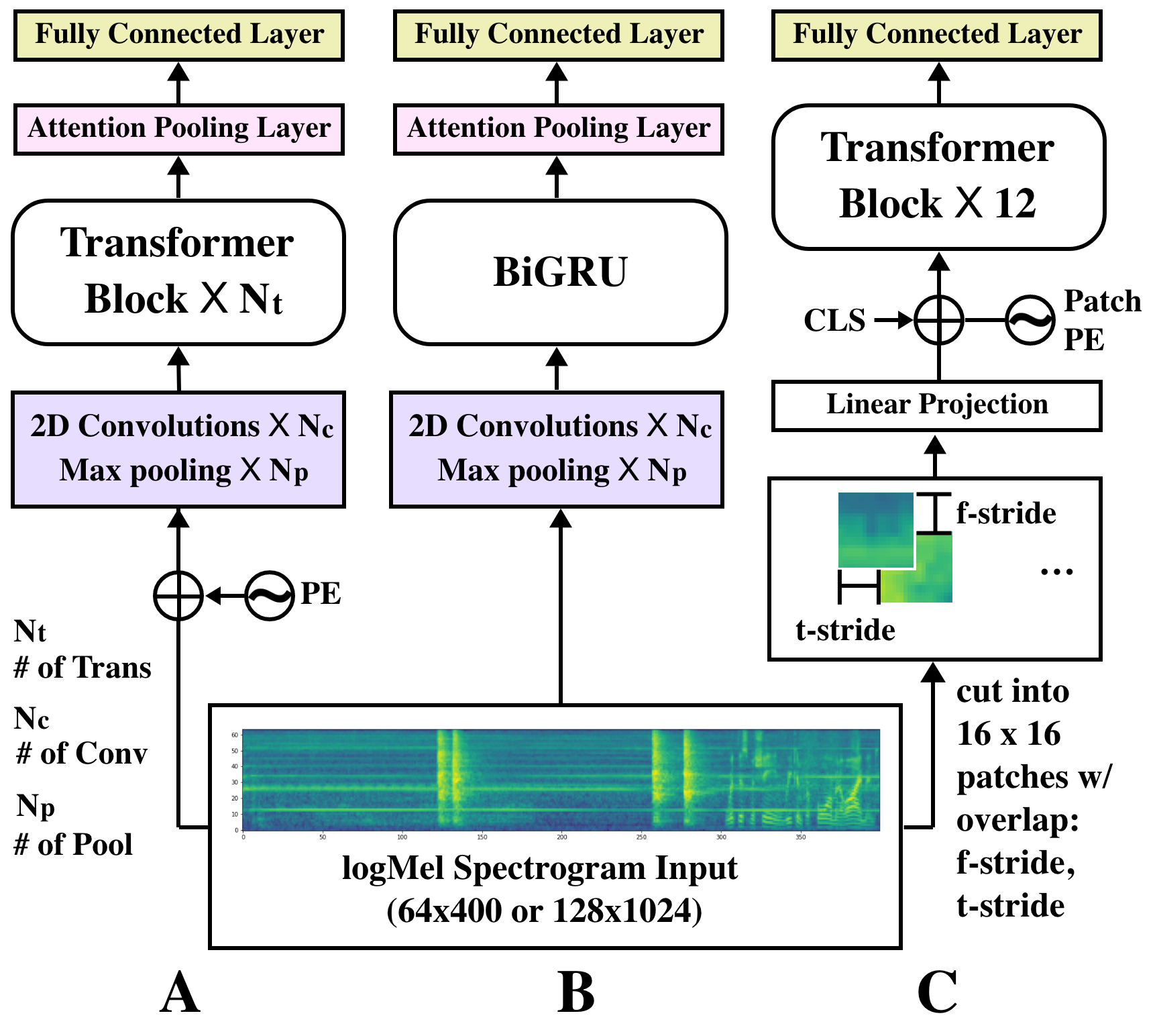}
    \caption{(A): CNN+Transformer, (B): CRNN (C): ViT. PE: positional encoding; $N_t=2$, $N_c=10$, $N_p=5$, f-stride$=$t-stride$=8$ }
    \label{fig:architecture}
\end{figure}
In this work, we focus on \emph{model robustness} without adopting any preprocessing step, and we focus on the Acoustic Event Detection (AED) task in the scope of this paper: Given an audio stream, we need to predict the most proper semantic label for the acoustic event.
General environmental sounds, compared to their well-studied human speech subset, do not usually contain strong contextual dependencies or senses, but span a wide range of frequencies and are naturally noisy. From a pure machine learning modeling perspective, we believe the AED task is a better angle for us to gain deeper understanding of specific neural architectures without worrying much about language models. 
Meanwhile, our findings about different neural models on the AED task could inform modeling choices for all other audio-related tasks such as ASR, speaker identification, and etc.

In AED task, CNN variants~\cite{kong2019panns} had many proven successes until being challenged by ViT~\cite{gong2021ast} architecture in 2021.
A recent comparison~\cite{li2022ATbaseline} showed CNNs are more efficient and easier to train compared to ViT, whereas ViT can beat pure CNNs in performance (mean Average Precision -- mAP) by a small margin on the AED task.
Despite the neck-to-neck performance, fundamentally, these 2 types of neural models are largely different in inductive biases due to their disparate learning mechanisms: convolutional neural networks are efficient since they effectively downsample the input with a sliding kernel (learned), discretizing the mathematical convolution into polynomial multiplication; whereas ViT networks model the global correlation among patches through the self-attention mechanism, which is inherently quadratic.
Given such distinctions, we suspect the 2 types of networks would have very different properties (e.g. robustness) when performing audio classification. Some recent work in the vision domain advocated for ViT's robustness~\cite{paul2021vision}, while its robustness on audio recognition (AR) remain unexplored. 
\begin{figure}[!ht]
    \includegraphics[width=1.0\linewidth]{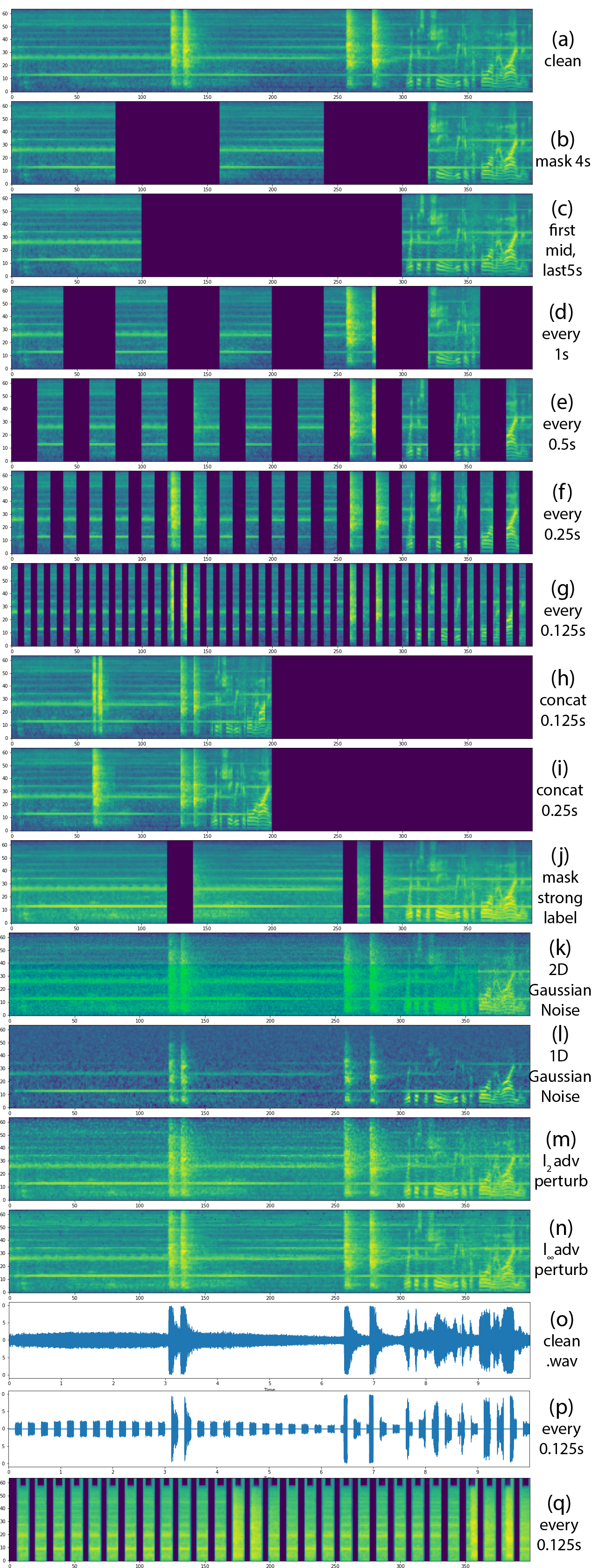}
    \caption{Different masking of the 10s Audio. The sample sound here is a ``hammer" class. Youtube id: --BfvyPmVMo. The 4 verticle lines are the actual hammering sound. }
    \label{fig:occlusion}
\end{figure}
\todo{F: I think we could rename "mask 4s" to "every 2s", and change the ordering of the presentation in the figure and the table? Maybe we could also have a condition where we cut one third from the middle (everything else, except "every 2s" cuts half), but the results are probably predictable.}

In this work, we seek to thoroughly understand the difference of noise robustness between the CNN variants and ViT models on the AED task. We pick competitive single-checkpoint models from the recent study~\cite{li2022ATbaseline}, and select 4 variants of CNN and transformer networks for comparison.
Our comparison includes CNN+Transformer, Vision Transformer (ViT), ResNet and CRNN on the task of Audio Event Detection (AED) using the largest dataset available: AudioSet~\cite{gemmeke2017audio}.

In our study, we start with the model's ``clean'' performance on the AudioSet test set without any input perturbation, and report performance changes when the model is under the influence of various type of noises. 
We first mask out a portion of the audio to test the change of the performance of different models.
Then, we model intermittent noise through \emph{occlusion}, and we model continuous noise using \emph{Gaussian} noise. 
In addition, we probe the model by computing \emph{adversarial examples} to understand the ``worst" case scenarios.
Our contributions:
\begin{enumerate}
    \item We systematically compared the performances of different transformer variants between several CNN variants under the influence of different input noises on the large-scale dataset Audioset.
    Our observations suggest that attention models bear very different \emph{model-based inductive bias} than pure convolution counterparts.
    \item We measure robustness of these different neural architectures, and we conclude that, trained on the same datasets doing the same task, self-attention architectures are more robust to noise noise perturbations than convolutional models in all regards.
\end{enumerate}

\section{Acoustic Event Detection Baselines}
\noindent \textbf{AudioSet~\cite{gemmeke2017audio}} contains 2,042,985 10-second YouTube video clips, summing up to 5,800 hours annotated with 527 types of sound events (weak label\footnote{does not specify which second specific event happens}). The same group~\cite{hershey2021benefit} released a set of \emph{temporally-strong} labels (103,464 train, 16,996 eval).
The \emph{full} trainset has 2 subsets: class-wise \emph{balanced} set (22,176 samples) and \emph{unbalanced} (2,042,985 samples) set, and \emph{eval} set with 20,383 samples~\cite{vggish}. \\
\noindent \textbf{Benchmark on the AED task:}
Table~\ref{tab:model-perform} lists state-of-the-art \emph{single models} and for the AED task trained on the \emph{full} AudioSet and test on the \emph{eval} set.\footnote{The waveform is downsampled to 16 kHz; frames of 1,024 samples (64 ms) are taken with a hop of 400 samples (25 ms); each frame is Hanning windowed and padded to 4,096 samples before taking the Fourier transform; the filterbank of 64 triangle filters spans a frequency range from 0 Hz to 8 kHz.}
To limit variability, we do not include (implicit \& explicit) ensembled models, multi-modal models. In~\cite{li2022ATbaseline}, they trained 6 model types, we select 4 models with competitive performance, so that comparison could be carried out fairly. 
The training procedure details and parameters are identical to what is described in~\cite{li2022ATbaseline}.
The architecture of each model are shown in Figure~\ref{fig:architecture}, ResNet50's architecture is omitted due to its prevalence.\todo{F: is this a good reason? I would add it, if we can. I understand the Figure may need to become bigger, but it is confusing later on, e.g. in Table 2, when we have A B C but nothing for ResNet...}
We use the 64x400 feature input models in the experiments of this work for better efficiency. (Shown in bold in Table~\ref{tab:model-perform})
\todo{F: So, in the table, we list two different sizes, but in this paper we only use one? Why not remove 128x1024 entirely then?}
\begin{table}[ht]
\setlength\tabcolsep{3.2pt}
\begin{center}
\caption{The best mAP of various models in comparison, w/. or w/o ImageNet pretraining, and using 64$\times$400 logMel spectrogram features.}.
\begin{tabular}{c c| c c  c c  } 
\toprule
\multirow{2}{*}{\textbf Model} & \multirow{2}{*}{\textbf \#param} & \multicolumn{2}{c}{64$\times$400}\\
&& {w/ pretrain} & {w/o pretrain}      \\
\midrule
AST/ViT &87.9M &\textbf{0.410} & 0.264 \\
CNN+Trans & 12.1M  & - & \textbf{0.411}\\
ResNet50 &25.6M & \textbf{0.382} &  0.370\\
CRNN  &10.5M  & - & \textbf{0.406}\\
\bottomrule
\end{tabular}
\label{tab:model-perform}
\end{center}
\end{table}
\todo{F: can we clarify if these numbers are the ones from the papers (which ones? let's add the references here), or are they the ones from your re-implementation? This needs to be very clear. How much effort would it be to run the ResNet-50 experiments with 128x1024 resolution? If we did, the baseline performance for ResNet would be the exact same as for the other architectures. It may be good to briefly describe the training of the various models here. Which one uses augmentation, dropout, other techniques etc that may interfere with the robustness?}

\section{Experiments \& Results}
\todo{F: We could briefly explain why we pick these four cases - or have we done it before, and I missed it?}

\subsection{Natural Occlusions}
To verify our hypothesis that attention models and convolution models are fundamentally different, we leverage time-domain occlusion as probing tool for analysis.
We apply blank (silent) temporal masks to cover different parts of the 10s audio, and apply it to all the samples in the \emph{eval set} of AudioSet, as shown in Figure~\ref{fig:occlusion}(a-g).
Then, we measure the performance change on the \emph{eval set} of our trained model.
Note that in Figure~\ref{fig:occlusion}(o,p), if we were to mask WAV file with silences and then extract logMel feature from the masked WAV file, we would end up with \href{https://en.wikipedia.org/wiki/Gibbs_phenomenon}{Gibbs phenomenon} due to performing FFT over the step function, namely the occlusion. 
In this paper, since our focus is to study the neural models' response over occlusion, the overshooting thorn-shaped artifacts (Figure~\ref{fig:occlusion}(q)) and the narrower occlusion on spectrogram (caused by FFT window leaking information into silence) are not desire.
Therefore, we occlude logMel features directly without worrying about such behaviors.\todo{F: Can we not just eliminate this discussion about masking raw audio files? We mask logMel spectrograms, for all models. End of discussion. Saves space. Billy: I will move it to appendix, for anyone who might be interested, since it is an interesting phenomenon} 

As we can see in Table~\ref{tab:audioOcclusion}, intuitively, the less information the model sees, the lower performance it would lead to.
This can be observed at Figure~\ref{fig:occlusion}(a-c), Table~\ref{tab:audioOcclusion} first 5 rows show that occluding consecutive blocks of sound by 40\% (i.e. mask 4s out of 10s) and 50\% would degrade the performance of all the 4 models by 6\% and 8\% mAP respectively. \todo{F: I think we can simplify the table here, and only report the numbers for "Mid 5s" occlusion, and say that we verified, and the result for first and last occlusion are very similar - should save some space}

\todo{F: I think 'consecutive' and 'intermittent' masking are really the same thing, except that for 'slow' change intervals, there can be 'rounding errors' and we retain less than half of the info. But we should simplify the presentation, and I am also not sure if we need to formally define the making function?}
On top of consecutive masking, we experimented with intermittent masking, which can be denoted as:
\newcommand*{\Cases}{\begin{cases} 1 & \text{if} \quad t \in [(n-1)d, nd], n\in\{2,4,6,...\frac{T}{d}\} \\ 0 & \text{if} \quad  \text{else}
\end{cases}
}%
\newcommand*{\VPhantom}{\vphantom{\Cases}}%
\newcommand*{\VPhantomHack}{\vphantom{\Cases^{2}}}%
\newcommand*{\BoldLeftBrace}{\pmb{\left\{\VPhantomHack\right.}}%
\newcommand*{\RightBrace}{\left.\VPhantom\right\}}%
\begin{equation}
    \mathbbm{1}_{t}(X_t) 
        = \Cases
\end{equation}
where $d$ is the masking interval, and $T$ being the total length of audio, $T=10s$ for AudioSet samples.
As Table~\ref{tab:audioOcclusion} row 6-9 show, the shorter the mask intervals, the lower the performance would be, despite the fact that the equivalent information is the same between masking every 0.125s and masking every 1s. 

Surprising, we could observe the existence of a threshold masking interval which significantly decreases the model performance. For instance, masking $d=0.25s$ for AST/ViT and CRNN almost nullified the models capability to recognize. 
As George Miller's experiment~\cite{cowan2015george}, neural networks resemble human memory in some ways, showing there's a ``magic number" period of information machine perception could remember.
Taking a closer look into the class-wise performance, under $0.125s-0.25s$ masking, there are only a handful of classes that the models could still predict with above 0.3 mAP: Speech, Music, and Whistling, and whispering, smoke detector and fire alarm. 
Even the classes previously found~\cite{li2019adversarial} to be robust to high-frequency adversarial noise, e.g. Police car, emergency vehicle, and siren classes are beyond recognizing for our model. 

Counter-intuitively, if we assemble the $d=0.125s$ fragments which were not masked to form a new 5s audio, the performance would recover dramatically, despite that we can clearly see the cutoff effect from the spectrograms. (see Figure~\ref{fig:occlusion}(h,i), Table~\ref{tab:audioOcclusion} row 10-13)
Interestingly, the performance of reassembled spectrogram would ``recover" to the equivalent value as if the mask was a consecutive 5s mask on the spectrogram, comparable to Table~\ref{tab:audioOcclusion} row 2-5.
Several observations:
\begin{enumerate}
    \item ViT/AST architectures are comparatively more robust architecture under occlusion, with its performance comparatively outperforms all other models under different type of occlusions.
    CRNN's robustness is also very competitive.
    \item If noise spans longer than the convolution receptive field, it causes huge performance drop, this specific behavior could be leveraged to infer neural architecture under a blackbox setting.
    \item This is a very effective and novel and super potent threat model for adversarial attack in audio domain. The above finds are supposed to work mainly on 2D FFT features.
\end{enumerate}
\todo{F: It is indeed weird that the model performance does recover so much for the "concat" case, but there is really no difference in the performance of the different methods, right? They all recover and behave quite the same. How big is the convolution receptive field for ResNet and CRNN? For the second observation, it would be important to show that it is comparable to where the dropoff occurs. How is all of this a super potent threat model?}

\begin{table*}[ht]
\setlength\tabcolsep{3.2pt}
\begin{center}
\begin{tabular}{c| c c c | c c c | c c c | c c c } 
\toprule
\multirow{2}{*}{\textbf Model} &  \multicolumn{3}{c|}{\textbf AST/ViT} & \multicolumn{3}{c|}{\textbf CNN+Trans}& \multicolumn{3}{c|}{\textbf ResNet} & \multicolumn{3}{c}{\textbf CRNN}\\
& { mAP} & {AUC} & {d-prime}  &  {mAP} & {AUC} & {d-prime} & {mAP} & {AUC} & {d-prime} & {mAP} & {AUC} & {d-prime}\\
\midrule
No Occlusion          & 0.411 & 0.837 & 1.387&0.411 & 0.971 & 2.672 & 0.382 & 0.966 & 2.574 & 0.406 & 0.972 & 2.694\\ \hline
4s Occlusion          & 0.360 & 0.798 & 1.178&0.349 & 0.956 & 2.417 &0.313 & 0.951 & 2.347 & 0.357 & 0.962 & 2.513  \\
First 5s Occlusion    &  0.321 & 0.748 & 0.946 &0.315 & 0.939 & 2.187  &  0.296 & 0.940 & 2.201 &0.326 & 0.947 & 2.286 \\
Mid 5s Occlusion      & 0.337 & 0.772 & 1.056&0.331 & 0.949 & 2.313   & 0.299 & 0.947 & 2.284 & 0.339 & 0.955 & 2.404 \\
Last 5s Occlusion     & 0.331 & 0.756 & 0.981&0.323 & 0.940 & 2.196  &  0.301 & 0.940 & 2.195 &0.329 & 0.945 & 2.266 \\\hdashline
Every 1s Occlusion    & 0.321 & 0.767 & 1.033 &0.292 & 0.941 & 2.214&0.237 & 0.929 & 2.079 &0.308 & 0.950 & 2.331 \\
Every 0.5s Occlusion  & 0.236 & 0.686& 0.686&\textbf{0.140} & 0.881 & 1.667  & \textbf{0.130} & 0.885 & 1.696 &0.213 & 0.925 & 2.037 \\
Every 0.25s Occlusion & \textbf{0.052} & 0.279 &-0.829&0.033 & 0.729 & 0.861  &0.020 & 0.637 & 0.494 & \textbf{0.067} & 0.821 & 1.299\\
Every 0.125s Occlusion&0.032 & 0.187 &-1.258&0.015 & 0.634 & 0.486 & 0.014 & 0.609 & 0.391 & 0.018 & 0.666 & 0.608 \\\hdashline
0.125s Concat         & 0.293 & 0.747& 0.939&0.281 & 0.941 & 2.209 & 0.266 & 0.940 & 2.199 &0.296 & 0.949 & 2.315  \\
0.25s Concat          & 0.313 & 0.766& 1.024&0.305 & 0.947 & 2.292 & 0.277 & 0.943 & 2.236 &0.314 & 0.954 & 2.385 \\
0.5s Concat           &0.333 & 0.781& 1.095 &0.326 & 0.951 & 2.338&0.294 & 0.947 & 2.284 & 0.334 & 0.957 & 2.422 \\
1s Concat             &0.340 & 0.784 & 1.109 &0.333 & 0.951 & 2.344&0.302 & 0.948 & 2.300 & 0.339 & 0.957 & 2.423 \\\hline
White Noise 2D& 0.248 & 0.694 & 0.718 &0.056 & 0.727 & 0.854 & 0.165 & 0.881 & 1.672 & 0.049 & 0.738 & 0.903 \\
While Noise 1D& 0.191 & 0.572 &0.258 & 0.155 & 0.827 & 1.335&0.147 & 0.815 & 1.265 &0.158&0.817&1.277\\\hline
Masking Strong & 0.203 & 0.495 &-0.019 & 0.194 & 0.792 & 1.149 &   0.183 & 0.826 & 1.328  & 0.193 & 0.830 & 1.347\\\hline
$l_{\infty}$ attack &0.260 & 0.700 &0.742 &0.164 & 0.891& 1.739& 0.011 & 0.542 & 0.148 & 0.082 & 0.845 & 1.434  \\ 
$l_{2}$ attack &  0.263 & 0.706 & 0.768 & 0.117 & 0.856 & 1.502 & 0.025 & 0.608 & 0.386  & 0.144 & 0.897 & 1.786 \\
\bottomrule
\end{tabular}
\end{center}
\caption{Different architectures(Audio Only) under different temporal occlusions}
\label{tab:audioOcclusion}
\end{table*}
\todo{F: why are some numbers bold? we do not need to repeat that this is audio-only here, right? would it not make sense to reverse the order of the "concat" lines, to be in sync with the "occlusion" ones? It would also be interesting to report the "50\% duration", the length of occlusion that halves the model's performance. For ResNet, this may be about 0.75s, for CRNN and AST/ViT this may be 0.375s. What does this tell us about the robustness of the various models? Somehow it says that AST/ ViT does not need more temporal context, but less, right?}

\subsection{Strong label masking}
With the release of new strong labels from AudioSet\cite{hershey2021benefit}, we have knowledge of when does specific event happen exactly for 16,996 samples among the entire 20,123 \emph{eval set}.
We mask out the specific strongly labeled portion of each audio sample for the entire eval set of AudioSet, and compare different model's test performance on the audio without the labeled events.
In essence, we are either occluding the most relevant part as shown in Figure~\ref{fig:occlusion}(j).
The results are shown in table~\ref{tab:audioOcclusion} on the line ``Masking Strong".
As we can see, it is surprising that the models are still preserving average 0.19 mAP, indicating the neural models are leveraging a tremendous portion of context around the labeled event in the audio, to predict the actual audio event in this case. ViT/AST models comparatively ``remember" the most information.
\todo{F: I think the most interesting observation is that all the models preserve about the same amount of information. What results do you get when you test only on the strong labeled test data? We have established that concatenating "cut" segments does little damage, so do all the models recover entirely on the "strong" labels?}


\subsection{Gaussian Noise Masking}
It is a common practice to add Gaussian noise to speech to model continuous noise. We test 2 types of Gaussian noises in this work: 
1) We add Gaussian noise $\sim \mathcal{N}(0,0.1)$ to the .wav file and let the noise cascade into the logMel filters through the FFT step and the Mel Filters.
2) We directly add Gaussian noise $\sim \mathcal{N}(0,0.1)$ onto the logMel spectrogram feature. 
Both Guassian noises are added to all the samples in the eval set of AudioSet, on which we report performances of different models in comparison.
Figure\ref{fig:occlusion}(k,l) illustrate the resulting logMel spectrograms of these 2 types of Gaussian noises. 
As we can see, the resulting spectrogram of 1D Gaussian noise is visibly lower magnitude than the 2D Gaussian noise.\todo{F: Presumably, 1D noise is the one directly applied to the audio?}
Table~\ref{tab:audioOcclusion} showcase the resulting differences. 
There are some surprising observation here, ViTs are more susceptible to 1D Gaussian noise whereas CRNN and CNN+Transformers are more susceptible to 2D Gaussian. 
This is potentially due to the attention mechanism of ViT are heavily leveraging temporal relations which would get more influence from the 1D Guassian noise. 
\todo{F: AST still does better in 1D noise than any other method, but it is striking that CNN+Trans and CRNN completely break down for 2D noise. Why is that? ResNet also is a convolutional architecture, but it seems to be considerably more robust. Also, the results of the occlusion study do not indicate that AST/ ViT heavily leverages temporal relations, because the model is more robust under frequent occlusions, or are you thinking at a entirely different time scale (samples)?}


\subsection{Adversarial Masking}
Finally, we want to study the ``worst case" scenarios by computing adversarial perturbations against the models.

\subsubsection{Adversarial Perturbations background}
%
The problem of computing universal adversarial perturbation to attack a classification model $f$ by maximizing the following~\cite{madry2017towards}:
\begin{equation}
\begin{aligned}
&
  \mathbf{E}_{(x,y)\sim \mathcal{D}} \underset{x' \in C(x)}{\max   }[L(f(x'),y)]\\
& \text{subject to  }
C(x)=\{ a\in \mathbb{R} :||a-x||_p \leq \epsilon \}.
 \\
\end{aligned}
\end{equation}

\noindent where $L$ is the loss function, $x$ is input and $y$ is label, $\mathcal{D}$ is the dataset, and $x' = x + \delta$ is our perturbed input. We want to find some perturbation $x'$ that looks ``indistinguishable'' from $x$, yet is classified incorrectly by $f$ even when $x$ is classified correctly.
To solve such a constrained optimization problem, one of the most common methods utilized to circumvent the non-exact-solution issue is the Projected Gradient Descent (PGD) method~\cite{madry2017towards}:
\begin{equation}
    \delta := \mathcal{P}_{\epsilon} \left(\delta - \alpha \frac{\nabla_{\delta}L(f( x +\delta),y)}{\Vert \nabla_{\delta}L(f( x +\delta),y)\Vert_p}\right)
\end{equation}
where $\mathcal{P}_{\epsilon}$ is the projection onto the $\ell_p$ ball of radius $\epsilon$, and $\alpha$ is the gradient step size.
To practically implement gradient descent methods, we use very small step size and iterate it by $\epsilon / \alpha$ steps. 

Recently, the research community has examine the robustness of the ViTs.  
\cite{paul2021vision, shao2021adversarial,bhojanapalli2021understanding} have concurrently studied and claimed that ViT are more robust learning compared to their CNN counterparts.
\cite{shao2021adversarial} advocates adversarial robustness is correlated to transformer blocks.
Meanwhile, ViT are shown to be more vulnerable against patch attacks~\cite{gu2021vision}. 
However, existing works mostly examine such robustness on vision tasks, not much work has been done to understand ViT's behaviors on audio dataset. 
Given the unique characteristics or audio, we do not expect the robustness ViT showed on vision tasks could be copied to audio tasks.
\todo{F: This would best go into the introduction or related work?}

\subsubsection{Adversarial Perturbation on LogMel Spectrogram}
In this work, we compute white-box adversarial perturbations against the 4 models and add them on the input feature (logMel spectrogram) level.\todo{F: Remind me: this is a white-box (or whateverit is called) attack, in which we have access to the model, and compute the gradients through it, correct?}
Specifically, we compute perturbation in both $\ell_{\infty}$ and $\ell_2$ norm, with the $\ell_{\infty}$ attack being a single step FGSM. 
Here, for every model that is under attack, we compute perturbation by PGD and use 20 step, step size 0.01, and $\epsilon = 0.1$. 
Figure.~\ref{fig:occlusion}(m) shows the effect of an $\ell_2$ norm perturbation. We can see the effect resembles Guassian noise but more potent. 
Figure.~\ref{fig:occlusion}(n)(needs to zoom in to see clearly) illustrates the FGSM attack in $\ell_{\infty}$ norm, the perturbation is effectively creating sparse zero-valued pixels in the logMel spectrogram.
All of our experiment results are shown in Table~\ref{tab:audioOcclusion} last 2 rows.
Here, we can see that ViT/AST models are once again more robust than other models. 
Interestingly, its performance under $\ell_{\infty}$ or $\ell_2$ norm attack with the same attack budget do not seem to differ much. 
ResNet's performance cratered under the influence of adversarial perturbation, indicating that without modules specifically modeling temporal dependencies, it is difficult to defend adversarial perturbations.
Also note here CRNNs are comparatively less robust against adversarial perturbation, despite being very robust against occlusions. 
It is intriguing that different neural architectures would respond to attacks under $\ell_{\infty}$ or $\ell_2$ norms very differently. 
\todo{F: This is very interesting indeed. Would it make sense to re-run these eperiments with multiple values of $\epsilon$? It should then be possible to produce a graph similar to the one that I produced for different values of occlusion and it would show where the dropoff occurs, and one could express the robustness of the model in terms of the $\epsilon$ value that reduces its performance to 50\%?}

\section{Conclusion and On-going work}
ViT/AST architecture are comparatively more robust compared to all the other 3 types of neural architectures against various types of noises, although the mechanism is currently still unknown and being actively investigated.
All neural networks learn/memorize vast amount of contextual data to perform predictions, the specific labeled events only accounts for 50\% of the mAP\todo{F: mAP} in AudioSet's case. 

We still do not have a clear way to explain the reason behind the drastically different behaviors of different neural architectures against adversarial perturbations. This is currently under investigation. 
In the mean time, we observe AUC, d-prime of ViT models are drastically different than all other 3 models in comparison, which could throw doubt on its robustness, such variations are currently being studied. 
\todo{F: Would it make sense to repeat the attacks also against the strongest Transformer model (trained with all tricks, e.g. dropout, mixup, augmentation and whatever else is there), to see if they make the AST model even more robust than it is already?}

\bibliographystyle{IEEEtran}

\bibliography{egbib}


\end{document}